\documentclass[review]{elsarticle}
\usepackage{amsfonts,amssymb,amsmath,amsbsy,amsthm,amscd}
\usepackage[english]{babel}
\usepackage{graphicx}

\newtheorem{theorem}{Theorem}[section]

\newtheorem{lemma}[theorem]{Lemma}
\newtheorem{corollary}{Corollary}

\newtheorem{remark}{Remark}

\renewcommand{\le}{\leqslant}
\renewcommand{\ge}{\geqslant}

\begin{document}

\begin{frontmatter}

\title{Singularity formation for rotational gas dynamics }


\author[1]{Olga S. Rozanova}
\ead{rozanova@mech.math.msu.su} 

\address[1]{Department of Mechanics and Mathematrics, Moscow
State University, Moscow 119991 Russia}


\begin{abstract}
The Cauchy problem for the system of equations of two-dimensional rotational gas dynamics is considered. It is assumed that the Cauchy data are a smooth compact perturbation of a constant state. Integral conditions for the data sufficient for the loss of smoothness by a solution within a finite time  are found. We analyze the possibility of fulfilling these conditions and compare them with the criterion of singularity formation, known for rotational gas dynamics without pressure.
\end{abstract}


\def\sign{\mathop{\rm sgn}\nolimits}






\end{frontmatter}


\section{Introduction}
We consider a system for density $\varrho({\bf x},t)$, pressure
$p({\bf x},t)$, velocity ${\bf u}({\bf x},t)$ and entropy $S({\bf
x},t)$:
\begin{eqnarray}
 \partial_t (\varrho{\bf u}) +   {\rm div}_{\bf x} ({\varrho\bf
u}\otimes{\bf u}) +
 \rho l{\mathcal L}{\bf u}+ \nabla_{\bf x} p  &=&  0, \label{e1}\\
\partial_t \varrho +  {\rm div}_{\bf
x} (\varrho {\bf u}) &=& 0, \label{e2} \\
\partial_t S + {\bf u}\cdot\nabla_{\bf x} S&=&0,\label{e3}\\
p=\exp S \varrho^\gamma. \label{e4}
\end{eqnarray}


Here ${\bf x}\in{\mathbb R}^2,$ $t\ge 0$, $\gamma>1$
is the heat ratio, $\mathcal L =\left(\begin{array}{cr} 0 & -1 \\
1 & 0
\end{array}\right)$, $l={\rm const}\ge 0$ is  the Coriolis parameter.

The initial data are the following:
\begin{eqnarray}
({\bf u},\varrho, p)|_{t=0}=({\bf u}_0,\varrho_0, p_0)({\bf
x})\label{CD}\in C^1({\mathbb R}^2),\\ \quad (\varrho_0, p_0)({\bf
x})>0,\quad ({\bf u}_0,\varrho_0, p_0)({\bf x})=({\bf 0},
\bar\varrho_0, \bar
p_0) \, \mbox{for}\, {\bf x}\notin B_R(0), \nonumber\\
\bar\varrho_0, \bar p_0 ={\rm const}>0,\,B_R(0)=\{{\bf x} \big||{\bf x}|<R,\,
R={\rm const}>0\}.\nonumber
\end{eqnarray}

The system is important because of geophysical applications, since it describes the height-averaged air movement in the atmosphere for middle-scale processes \cite{Pedlosky}.  For $\gamma=2$ the system \eqref{e1} -- \eqref{e4} corresponds to the equations of rotating shallow water. If $ l = 0 $, then these are the standard hyperbolic compressible Euler equations for polytropic gas; a review of the properties can be found in \cite{Chen}. System \eqref{e1} -- \eqref{e4} can be written in a symmetric form, therefore the  solution of the Cauchy problem \eqref{e1} -- \eqref{CD} keep initial smoothness at least for small $t$ \cite{Kato}. At the same time, the solutions of nonlinear hyperbolic systems have the property of losing smoothness, therefore one of the interesting
and difficult problems is to find a class of initial data leading to
a blowup in a finite time.  In the well-known work \cite {Sideris}, integral conditions for the initial data were found that are sufficient for a loss of smoothness. This work has generated many results of this kind for gas dynamics, as well as for systems associated with it, see, for example, \cite {Rozanova_JMS} and references therein.  The results are in some ways simpler for compactly supported solutions
\cite{Xin}. The most elegant theorems regarding energy balance can be obtained for solutions with a finite moment of mass,
the pioneering work was \cite{Chemin}. Energy balance and sufficient conditions for a
 singularity formation for compactly supported solutions of \eqref{e1} -- \eqref{e4} were obtained in \cite{Rozanova_DU}, similar results for
 solution with a finite moment of mass are contained in
 \cite{Rozanova_JMS}. Estimates of unavailable potential energy were obtained in \cite{RozanovaFAO}.
 An important result demonstrating that rotation prevents the formation of a singularity can be found in \cite{Tadmor_Lui}, \cite{Cheng_Tadmor}.

 We note that the issue of the formation of a singularity is very important in the meteorological context since singularities are associated with  atmospheric fronts. In addition, knowledge of the  class of initial data leading to a blowup helps to study the possibility of the existence of large atmospheric vortices such as typhoons.

Let us introduce the following functionals.

\begin{eqnarray*}
  G(t) = \frac{1}{2}\int_{B_R(t)}|{\bf x}|^2 \varrho\, dV \ge 0, \\ F_1(t) = \int_{B_R(t)} \varrho {\bf u}\cdot {\bf x} \, dV,
 \quad F_2(t) = \int_{B_R(t)} \varrho {\bf u}\cdot {\bf x}_\bot \, dV,\quad {\bf x}_\bot =(x_2, -x_1) \\
 E_k(t)=  \frac{1}{2}\int_{B_R(t)} \varrho |{\bf u} |^2\, dV, \quad e(t) = E_k(t)+\frac{1}{\gamma-1} \int_{B_R(t)} (p- \bar p) \, dV,  \\
 \quad m(t) = \int_{B_R(t)} (\varrho-  \bar\varrho) \, dV,\quad
 {\bf P}(t)= \int_{B_R(t)} \varrho {\bf x}\, dV,\quad {\bf I}(t)= \int_{B_R(t)} \varrho {\bf v}\,
 dV.
\end{eqnarray*}

Such functionals are very convenient for studying various properties of a rotating gas (see \cite{Rozanova_DU}, \cite{Rozanova_JMS},
\cite{RT1},  \cite{RT2})

We introduce the notation: $B_R(t)=\{{\bf x}\big| |{\bf x}|<R+\sigma
t\}$, $\sigma=\sqrt{p_\varrho}|_{\varrho=\bar\varrho}$ is the sound
speed (speed of propagation of perturbations).

 First of all we note that for $C^1$ - smooth solutions of \eqref{e1} -- \eqref{CD} the support of perturbation is contained in  $B_R(t)$.
 It is a corollary of the local energy estimates and can be proved  as in \cite{Sideris1} for symmetric hyperbolic systems. The rotational term does not give any difference in the prove, since ${\bf u}\cdot {\mathcal L}{\bf u}=0$.

\begin{lemma} For $C^1$ - smooth solutions of \eqref{e1} -- \eqref{CD} the following properties hold:
\begin{eqnarray}
 m'(t)=0,\label{dm}\\
e'(t)=0,\label{de}\\
 G'(t)=F_1(t)+\pi\bar\varrho \sigma(R+\sigma t)^3,\label{dG}\\
 F_2'(t)= l F_1(t),\label{dF2}\\
F_1'(t)= 2 E_k(t)- 2 \int_{B_R(t)} (p- \bar p )\, dV -  l F_2(t),
\label{dF1}\\
{\bf P}'(t)={\bf I}(t),\label{dP}\\
{\bf I}''(t)+ l^2{\bf I}(t)=0.\label{dI}.
\end{eqnarray}
\end{lemma}

\proof The proof is a direct calculation of the derivatives of the integrals over the moving volume, taking into account \eqref {e1}, \eqref {e2} and the Stokes formula.

\begin{corollary}
\begin{eqnarray}
l G(t)-F_2(t)= l G(0)-F_2(0)+\frac{l\pi\bar\varrho\sigma}{4}((R+\sigma t)^4-R^4) \label{F2G}\nonumber  \\
G''(t)+l^2 G(t)= 2 (2-\gamma)E_k(t)+ A_0 + q(t),
 \label{ddG}\\
A_0=2 (\gamma-1)e(0) + l^2 G(0)-lF_2(0),  \label{A0}\nonumber\\
q(t)= \frac{l^2\pi\bar\varrho}{4}((R+\sigma
t)^4-R^4)+3\pi\bar\varrho\sigma (R+\sigma t)^2. \label{q}\nonumber
\end{eqnarray}
\end{corollary}

\proof The first property follows from \eqref{dG} and \eqref{dF2}. Then together with \eqref{dF1} and \eqref{de} we get \eqref{ddG}.


\section{Inequalities leading to a contradiction}

\subsection{Lower and upper bounds of $G(t)$}\label{S_G_bound}

In this section, we obtain general estimates that are independent of $\gamma$.


\begin{lemma}
 For non-trivial $C^1$ - smooth solutions of \eqref{e1} -- \eqref{CD}
 \begin{equation}G(t)>0;\nonumber \end{equation}
\begin{equation} \label{G-} G(t)\ge \phi_-(t)\equiv \frac{|{\bf P}(t)|^2}{m(0)+\bar\varrho \pi (R+\sigma
t)^2},\end{equation} where \begin{eqnarray*}|{\bf
P}(t)|^2=P^2_1(t)+P^2_2(t),\, P_i(t)=\int_0^t I_i(\tau) d\tau,
i=1,2,\\ I_1(t)=I_1(0)\cos lt + \frac{I_2(0)}{l}\sin lt ,\quad
I_2(t)=I_2(0)\cos lt + \frac{I_1(0)}{l}\sin lt .\end{eqnarray*}
\end{lemma}

\proof The first statement is evident. To prove the second, we note that from
the H\"older inequality we have $$|{\bf P}(t)|^2\le 2 G(t)\int_{B_R(t)}
\varrho \,dV.
$$
The explicit form of ${\bf P}(t)$ can be found from \eqref{dI} and
\eqref{dP}. $\square$

\remark If ${\bf P}(t)=0$ (for example, for axisymmetric initial data), the lower estimate \eqref{G-} is zero. Otherwise \eqref{G-}
is more exact.

\begin{lemma}
For non-trivial smooth solutions of  \eqref{e1} -- \eqref{CD}
\begin{equation}\label{G+}
 G(t)\le \frac{(R+\sigma t)^2}{2}(m(0)+{\bar\varrho
\pi} (R+\sigma t)^2)\equiv \phi_+(t).
\end{equation}
\end{lemma}

\proof The estimate follows from \eqref{dm} and the inequality
$$G(t)\le\frac{(R+\sigma t)^2}{2}\int_{B_R(t)}
\varrho \,dV. $$

\subsection{The case $\gamma=2$, the shallow water equations}

In this case equation \eqref{ddG} can be easily solved, namely,
\begin{eqnarray}\label{G2}
G(t)=\frac14\sigma^4\bar\varrho \pi t^4+\sigma^3\bar\varrho \pi R
t^3+\frac32 \sigma^2\bar\varrho \pi R^2 t^2 +\sigma \bar\varrho R^3
t+\\\frac{F_1(0)}{l} \sin lt+(G(0)-\frac{A_0}{l^2})\cos lt
+\frac{A_0}{l^2}.\nonumber
\end{eqnarray}
Thus, if the behavior of $G(t)$ for some $T$ contradicts the upper and lower estimates proved in Sec.\ref{S_G_bound},this means that the solution loses smoothness up to this point in time.

We obtain the following result.

\begin{theorem} \label{TG2} Suppose $\gamma=2$ and there  exists a positive $T_*$ such that the graphs of
functions $G(t)$, given by \eqref{G2}, intersects with  $\phi_-(t)$,
given by \eqref{G-}, or with $\phi_+(t)$, given by \eqref{G+}. Then
the classical solution of the Cauchy problem \eqref{e1} -- \eqref{CD} loses smoothness during $T<T_*$.
\end{theorem}

{\remark Since $ G (t) $ is the sum of increasing and oscillating functions, it is easy to see that if $ T_* $ exists, then
$T<\frac{2\pi}{l}$ and
\begin{equation}\label{Acond}
\left(\frac{A_0}{l^2}\right)^2-\left(\frac{F_1^2(0)}{l^2}+\left(G(0)-\frac{A_0}{l^2}\right)^2\right)\le
0.
\end{equation}
However condition \eqref{Acond} is not sufficient for the intersection
of graphs of $G(t)$ and $\phi_-(t)$.}

{\remark Condition \eqref{Acond} holds if and only if
\begin{equation*}\label{Acond_short}
-l^2 G^2(0)+2 F^2_1(0)\ge 0.
\end{equation*}
It implies that $F_1 $ (i.e. the radial part of velocity) is initially large enough.}

Indeed, \eqref{Acond} implies
\begin{equation}\label{Acond1}
\frac{8}{l^4}e^2(0)+(\frac{4}{l^3}(2lG(0)-F_2(0)))
e(t)+\frac{(l^2G^2(0)-F_2(0))^2+F^2_1(0)+F^2_2(0)}{l^2}\le 0,
\end{equation}
the left hand side is a quadratic polynomial with respect to $e(0)$.
Thus, if the determinant of this polynomial, $-l^2 G^2(0)+2 F^2_1(0)$,
is negative, \eqref{Acond1} never holds.


\subsection{The case $\gamma\ne 2$}

In this case we need an additional lemma.

\begin{lemma} \label{lemma_pp0} Assume
\begin{equation}S_0(x)=\ln p_0(x)\varrho^{-\gamma}_0(x)\ge \ln \bar
p\bar\varrho^{-\gamma}\equiv \bar S, \, x\in {\mathbb R}^2.\label{S0}
\end{equation}
Then
\begin{equation}
\int_{B_R(t)}(p-\bar p) \,dV\ge \gamma m(0)
{\bar\varrho}^{\gamma-1}{\exp \bar S}=\sigma^2 m(0).
\label{pp0}\nonumber
\end{equation}
\end{lemma}

\proof Denote $|B_R(t)|= \pi(R+\sigma t)^2$. First of all, we notice that \eqref{e3} and
\eqref{S0} imply $S(t,x)\ge \bar S$ for smooth solutions. Thus,
\begin{eqnarray*}
\int_{B_R(t)}(p-\bar p) \,dV =\int_{B_R(t)}p \,dV-\bar p |B_R(t)|=\\
\int_{B_R(t)}\rho^\gamma \exp {S} \,dV-\bar p |B_R(t)|=
 \exp {\bar S}(\int_{B_R(t)}\varrho^\gamma \,dV-\bar\varrho^\gamma
 |B_R(t)|)\ge\\\mbox{[Jensen's inequality]}\\
 \exp {\bar
 S}(|B_R(t)|^{1-\gamma}((m(t)+\bar\varrho|B_R(t)|)^\gamma-\bar\varrho^\gamma
 |B_R(t)|)=\\
\exp {\bar S}\bar\varrho^\gamma
|B_R(t)|\left(\left(\frac{m(t)}{\bar\varrho
|B_R(t)|}+1\right)^\gamma-1\right)\ge\\
\mbox{[Bernoulli inequality]}
\\
\exp {\bar S}\bar\varrho^\gamma |B_R(t)|\left(
 \left(1+ \frac{\gamma m(t)}{\bar\varrho
|B_R(t)|}\right)-1\right)= \gamma m(0)
{\bar\varrho}^{\gamma-1}{\exp \bar S}. 
\end{eqnarray*}
We apply the Bernoulli inequality to obtain
$$\left(\frac{m(t)}{\bar\varrho |B_R(t)|}+1\right)^\gamma\ge
1+ \frac{\gamma m(t)}{\bar\varrho |B_R(t)|},$$ since
 $m(t)+\bar\varrho |B_R(t)|\ge 0 $ and therefore $\frac{m(t)}{\bar\varrho
|B_R(t)|}\ge -1$. $\square$

{\remark Property \eqref{S0} always holds for isentropic motion $(S=\rm const)$.}

\subsubsection{ $\gamma> 2$}

From \eqref{ddG} taking into account the fact that $G(t)$ and
$E_k(t)$ are nonnegative we have
\begin{eqnarray}\label{G2+}
G(t)\le f_+(t)\equiv Q(t)+
\frac{A_0}{2} t^2 + G'(0)t +G(0),\nonumber\\
Q(t)=\int_0^t \int_0^\tau q(\lambda) d\lambda d\tau.
\label{Q}\nonumber
\end{eqnarray}
It is a rough estimate, we can use lower bound \eqref{G-} to get
$-l^2 G(t)\le -l^2 \phi_-(t).$

Taking into account \eqref{G+} and Lemma \ref{lemma_pp0} from
\eqref{ddG} we also have
\begin{eqnarray*}
G''(t)=-l^2 G(t)  + 2 e(t) +
\frac{2(\gamma-2)}{\gamma-1}\int_{B_R(t)}(p-\bar p) \, dV \\ +l^2
G(0)-l F_2(0)+ q(t) \ge q(t)-l^2 \psi_+(t)+A_1,\\
A_1\equiv 2 e(0) +l^2 G(0)-l
F_2(0)+\frac{2(\gamma-2)}{\gamma-1}\sigma^2 m(0).
\end{eqnarray*}
Thus,
\begin{eqnarray}\label{G2-+}
G(t)\ge f_-(t)\equiv Q_1(t)+
\frac{A_1}{2} t^2 + G'(0)t +G(0),\nonumber\\
Q_1(t)\equiv Q(t)=\int_0^t \int_0^\tau (q(\lambda)- l^2
\psi_+(\lambda) ) d\lambda d\tau.\nonumber
\end{eqnarray}

Thus, we obtain the theorem.

\begin{theorem}\label{TG2+} Suppose that $\gamma>2$ and condition \eqref{S0} is satisfied.  Assume also that for initial data \eqref{CD}
there  exists a positive $T_*$ such that the graphs of
functions $f_+(t)$ and $\phi_-(t)$ intersect or the graphs of
functions $f_-(t)$ and $\phi_+(t)$ intersect.  Then the classical
solution of the Cauchy problem \eqref{e1} -- \eqref{CD}
loses smoothness within a time $T<T_*$.
\end{theorem}

\subsection{$\gamma<2$}

Analogously to the previous subsection  we have from \eqref{ddG}
\begin{eqnarray*}
G''(t)\ge q(t)-l^2 \psi_+(t)+A_0,\\
G''(t)= -l^2 G(t)  + 2 e(t) +
\frac{2(\gamma-2)}{\gamma-1}\int_{B_R(t)}(p-\bar p) \, dV \\ +l^2
G(0)-l F_2(0)+ q(t) \le q(t)+A_1,\\
G(t)\ge g_-(t)\equiv Q(t)+
\frac{A_1}{2} t^2 + G'(0)t +G(0),\nonumber\\
G(t)\le g_+(t)\equiv Q_1(t)+ \frac{A_0}{2} t^2 + G'(0)t
+G(0).\nonumber\\
\end{eqnarray*}

Thus, we obtain the theorem.

\begin{theorem}\label{TG2-} Suppose that $\gamma<2$ and condition \eqref{S0} holds.   Assume that for initial data \eqref{CD}
there  exists  a positive $T_*$ such that the graphs of
functions $g_+(t)$ and $\phi_-(t)$ intersect or the graphs of
functions $g_-(t)$ and $\phi_+(t)$ intersect.  Then a classical
solution to the Cauchy problem \eqref{e1}-\eqref{e4}, \eqref{CD}
loses smoothness within a time $T<T_*$.
\end{theorem}

{\remark If the time $T_*$ from Theorems \ref{TG2+} and \ref{TG2-}
exist, it is sufficiently small.}

\section{Analysis of sufficient conditions for blowup and examples}

In this section we are going to discuss the following questions: \begin{itemize}
\item Can sufficient conditions be satisfied for any data?
\item If so, is it possible to judge what kind of singularity arises?
\item How rough are the sufficient conditions for the singularity
formation? How far are they from the criterion?
\item What factors promote or prevent blowup?
\end{itemize}

1. Let us show that the first question is not trivial. Indeed, we can
set for the sake of simplicity $l=0,$ $\bar\varrho=\bar p=\sigma=0$
and consider axisymmetric initial conditions to have $\phi_-=0$.   Then for smooth
nontrivial solutions we have
\begin{eqnarray*}G''(t)=2e(0), \quad G'(t)=F_1(t),\label{GF}\\ e(0)=E_k(t)+\int_{B_R(t)}
p\,dV> E_k(t)>0.\nonumber
\end{eqnarray*}
Thus,
\begin{eqnarray}G(t)=e(0)t^2+G'(0)t+G(0),\label{GF}
\end{eqnarray}
and $G(t)$ vanishes if and only if
\begin{eqnarray}\label{sc}(G'(0))^2\ge 4 e(0)G(t).
\end{eqnarray}
Nevertheless, due to the H\"older inequality
$(G'(t))^2=(F_1(t))^2\le 4 E_k(t)G(t)< 4 e(0)G(t)$. Thus, we get a contradiction with \eqref {sc} and cannot find the initial data that lead to the blowup. At the same time, it is well known \cite{Xin} that any
solution with compactly supported initial data loses smoothness in a finite
time.

This example shows that sufficient conditions for vanishing $G(t)$
do not detect  the possible formation of a singularity.

 For our simple example, a different method may be proposed.
Namely, \eqref{GF} implies that $G(t)$ is unbounded. However, from
\eqref{G+} we get that in our case  $G(t)\le \frac{R^2}{2}
m(0)\equiv G_m$. This contradiction show that any nontrivial
solution blowups (in fact, we reproduce the method of proof from
\cite{Xin}).

2. In contrast to the case $l=0$,  for $l\ne 0$  sufficient
conditions from Theorems \ref{TG2}, \ref{TG2+}, \ref{TG2-}, not
always imply a blowup for compactly supported initial data.

For example, for any $\gamma>1$ we can write
\begin{equation*}\label{Gs} G''(t)\ge 2 e(0)- l^2 (G_m-G(0)) -l F_2(0)\equiv A_3.
\end{equation*}
Thus, to obtain a contradiction with the inequality $G(t)\le G_m$, we
can require, for example, $A_3>0$. Nevertheless, since $G_m-G(0)>0$,
for any initial data we can obtain $A_3<0$, increasing $l$.

3.  Analysis of sufficient conditions, considered in this paper,
show that as $l$ increases, then harder to detect a blowup. Of course
this still does not mean that the rotation prevents a blowup. We can draw this conclusion only if there
is a criterion for the formation of a singularity (see Example 2 for the case of pressureless gas
dynamics). Thus,  here we can conclude that increasing rotation we can obtain
globally smooth in time solution starting from any smooth initial
data.

Further, the absence of axial symmetry also promotes the implementation of sufficient conditions,
since a nonzero lower bound $\psi_-$ for $G(t)$  arises.

And finally, the lower the speed of sound (that is, the speed of propagation of the support), the simpler the fulfillment of the integral sufficient conditions.


4. From the H\"older inequality we have
\begin{equation}\label{Holder}
(F_1(t))^2 = (G'(t))\le 4 G(t) E_k(t),
\end{equation}
therefore if $G(t)\to 0$ as $t\to T$, and $F_1(T)\ne 0$, then $ E_k(t)\to \infty$ as
$t\to T$. Therefore, provided the solution keeps smoothness up to
the time $T$, the density and/or velocity tend to infinity.

5. Inequality \eqref{Holder} also allows to estimate the kinetic energy $E_k$ from below. Namely,
\begin{equation*}\label{Holder}
 E_k(t)\ge \frac{(F_1(t))^2}{  4 G(t)}\ge \frac{(F_1(t))^2}{  4 \psi_+(t)}.
\end{equation*}

6. The sufficient conditions considered here help to find perturbations of initial data which are so strong they almost immediately generate a singularity. Indeed, for large $t$ function $G(t)$ increases. However, the growth rate as $t\to\infty$ is slower than the growth rate of its upper boundary $\psi_+$ (for the case $\gamma=2$ it is easier to see).

\subsection{Example 1}

As an example, we consider perturbations of a steady vortex,
constructed following to \cite{Rozanova_frozen} in the isentropic
case ($S=\rm const$). Below we set $l=1$.

As the initial data, which are a perturbation of a constant state inside
$|{\bf x}|=r\le R=1$ we choose
\begin{eqnarray}\label{u0}
{\bf u}_0=\frac{\Phi' (r)}{r} \left(\begin{array}{cr} \epsilon  & 1 \\
-1 & \epsilon
\end{array}\right)\left(\begin{array}{c}  x_1 \\
x_2
\end{array}\right),
\end{eqnarray}
where
$$
\Phi=\frac{b}{4}\left\{\begin{array}{cr} r^2 (2-r^2),& r\le 1 \\
1, & r > 1
\end{array}\right.,
$$
and
$$p_0=  (\bar\Pi+\Pi)^\gamma,\quad
\varrho_0=\left(\frac{p_0}{C}\right)^\frac{1}{\gamma},$$ where
$$\Pi=\frac{b }{12}  \left\{\begin{array}{cr}   (2 b r^4+3 b (1-2b)r^2+6(b-1)),& r\le 1 \\
2 b-3,& r > 1
\end{array}\right..
$$

\begin{figure}[]
\center
{\includegraphics[width=0.5\linewidth]{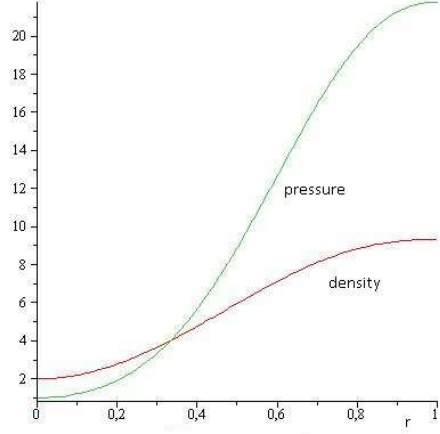}}
\caption{Profiles of density and pressure for a steady state solution
(Example 1)} \label{Pic1}
\end{figure}

We set $\gamma=2$, so we have an explicit formula for $G(t)$, and
choose $ C=\frac{1}{4}$, $\bar\Pi=1$. The constant $b$ is a measure
of vorticity, for our computations we choose $ b=-4.$ If
$\epsilon=0$, the initial data correspond to a steady state.
$\epsilon>0$ corresponds to initially divergent motion, whereas
$\epsilon<0$ corresponds to initially convergent one. Fig.1 presents initial profiles of density and pressure as functions of $r$
inside the support of perturbation.

\begin{figure}[h]
\begin{minipage}{0.5\columnwidth}
\centerline{\includegraphics[width=0.9\columnwidth]{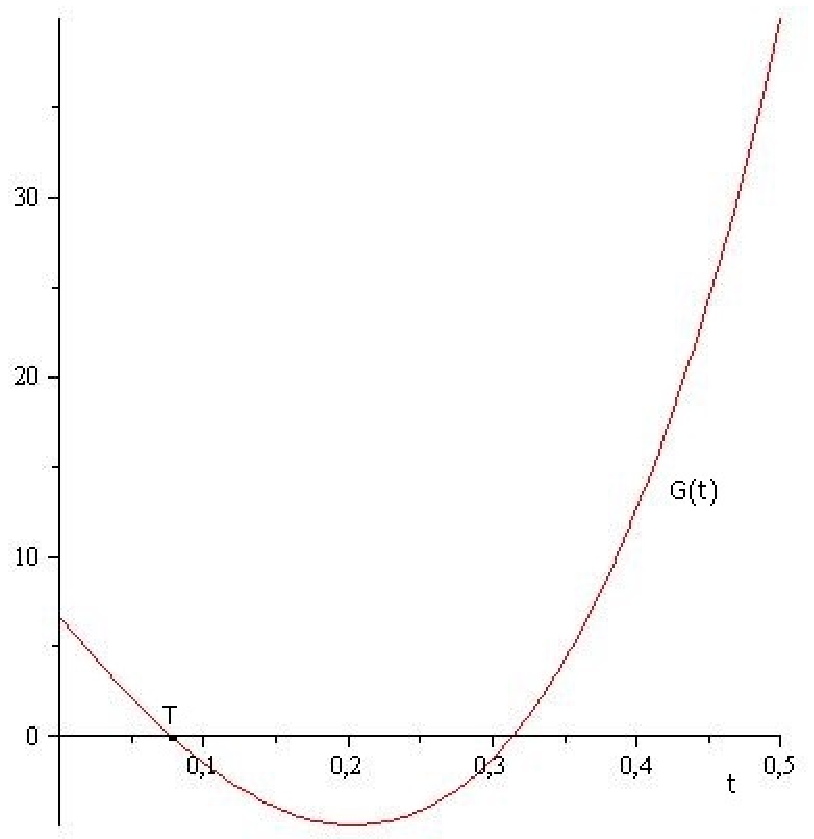}}
\end{minipage}
\begin{minipage}{0.5\columnwidth}
\centerline{\includegraphics[width=0.9\columnwidth]{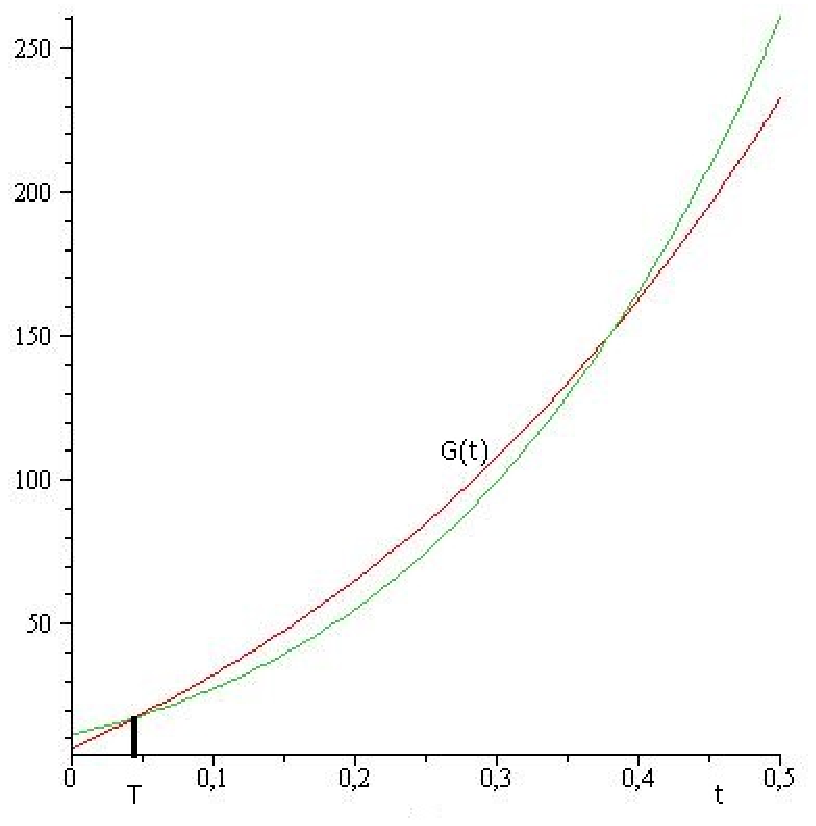}}
\end{minipage}
\caption{ $G(t)$ and $\phi_-(t)=0$ for $\epsilon=10$ (left);  $G(t)$
and $\phi_+(t)$ for $\epsilon=-10$ (right). }
\end{figure}

Pic.2 shows the intersection of graphs of $G(t)$ and $\phi_-=0$ (the
data are axisymmetric) for $\epsilon=10$, highly convergent motion
(left) and the intersection of graphs of $G(t)$ and $\phi_+(t)$
 for $\epsilon=-10$, highly divergent
motion (right). We can see that the singularity appears very fast.


\subsection{Example 2}

We study how far the sufficient conditions for the singularity formation
are far from the criterion on the example of pressureless gas
dynamics, i.e. $C=0$. It seems, that this the only example of multidimensional gas dynamics, where the criterion is known.
 It was
obtained in \cite{Tadmor_Lui} in Lagrangean formulation and in
\cite{Rozanova_Usp} in Eulerian formulation. In \cite{Rozanova_Usp}
an integral representation of the solution is also obtained.
Namely, a solution of
\eqref{e1}, \eqref{e2}, $p=0$, keeps smoothness for all $t>0$ if and
only if for every point $(x_1, x_2)\in {\mathbb R}^2$
\begin{equation}\label{ba1}
({\rm div}\, {\bf u}_0)^2-4\,J({\bf u}_0)-2 \,l \, {\rm rot}\, {\bf
u}_0-l^2<0,
 \end{equation}
where $J({\bf u}_0)={\rm det}\big(\frac{\partial u_{0i}}{\partial
x_j}\big)$, $i,j=1,2,$ ${\rm rot} {\bf
u}_0=(u_{02})_{x_1}-(u_{01})_{x_2}$.

As initial compactly supported velocity ${\bf u}_0$ we choose
\eqref{u0} and $\varrho_0=1,$ the parameter $l=1$ as before. It can
be readily checked by means of \eqref{ba1} that for $\epsilon=0$ the solution to the Cauchy
problem for \eqref{e1}-\eqref{e2}, $p=0$, keeps smoothness for all
$t>0$ if $b\in (-0.1,0.2)$ (left and right bounds are approximate).
We are going to test the solution with parameter $b=-0.05$.
 First we perturb the radial component of initial data. Computations
 show that the solution remain smooth only for $\epsilon \in (-7,2.5)$
If $\epsilon$ does not belong  to this domain there are points generating singularity (the bounds are approximate).
 For
 $\epsilon>0$ (a divergent motion)
 these points are close to the boundary of support,  for
 $\epsilon<0$ (a convergent motion)
 these points are close to the center.
The function $G(t)$ obeys the equation
$$G''(t)+l^2 G(t)=2 e(0) + l^2 G(0) - l F_2(0)\equiv A_4,$$
therefore the solution can be found explicitly. It is also easy to find conditions that contradict
 bilateral inequality $0\le G(t) \le G_m.$
The conditions look like \eqref{Acond}, namely
\begin{equation*}
\left(\frac{A_0}{l^2}\right)^2-\left(\frac{F_1^2(0)}{l^2}+\left(G(0)-\frac{A_0}{l^2}\right)^2\right)\le
0.
\end{equation*}
and
\begin{equation*}
\left(G_m-\frac{A_0}{l^2}\right)^2-\left(\frac{F_1^2(0)}{l^2}+\left(G(0)-\frac{A_0}{l^2}\right)^2\right)\le
0.
\end{equation*}
Both conditions give approximately the same limitations on
$\epsilon$ leading to a blowup, $|\epsilon |>30$. However, we saw that,
in fact, the solutions already  blow up  for $|\epsilon |>7$.
 Thus,
the sufficient conditions for a singularity formation considered here are
quite rough and the class of initial data such that the corresponding solution loses smoothness, much broader than that specified in integral sufficient conditions, so they are far from being precise.

\end{document}